\documentclass{article}
\usepackage[vlined,boxed,algonl]{algorithm2e}
\usepackage{algorithm2e-cgf}
\usepackage{epsfig}
\usepackage{url}
\usepackage[export]{adjustbox}
\usepackage{longtable}
\usepackage{graphicx, subfigure}
\usepackage{placeins}
\usepackage[table]{xcolor}
\usepackage{amsmath} 
\usepackage{amsfonts}
\newcommand{\Z}{\mathbb{Z}}

\title{Speeding Up Elliptic Curve Multiplication with Mixed-base Representation for Applications to SIDH Ciphers}
\author{Wesam Eid and Marius Silaghi\\Florida Institute of Technology}
\date{}

\begin{document}
\maketitle

\begin{abstract}
Elliptic curve multiplications can be improved by replacing the standard ladder algorithm's base 2 representation of the scalar multiplicand, with mixed-base representations with power-of-2 bases, processing the $n$ bits of the current digit in one optimized step. For this purpose, we also present a new methodology to compute, for Weierstrass form elliptic curves in the affine plane, operations of the type $mP+nQ$ where $m$ and $n$ are small integers. This provides implementations with the lower cost than previous algorithms, using only one inversion. In particular, the proposed techniques enable more opportunities for optimizing computations, leading to an important speed-up for applications based on elliptic curves, including the post-quantum cryptosystem Super Singular Isogeny Diffie Hellman~(SIDH). 
\end{abstract}

\section{Introduction}
  
The most popular forms of public-key cryptography for current applications have increasingly
been based on Elliptic Curves (ECs)~\cite{miller1985use,koblitz1987elliptic}. With Elliptic Curve Cryptography (ECC), messages and secrets are mapped to points on an elliptic curve, and specific point doubling and point addition operations define transitions between points. Scalar point multiplication uses a sequence of point doubling and point additions to efficiently evaluate point multiplications:
$$Q = [k]P = \underbrace{P + P + ... + P}_k.$$
Cryptosystems based on ECs rely on the difficulty of solving the Elliptic Curve Discrete Log (ECDL) problem. Namely, given the points Q and P in the previous equation,
 it is hard to determine the scalar multiple $k$ for elliptic curves with points $P$ of large order and large $k$ numbers. However, with the expected emergence of quantum 
computers~\cite{chen2016report} in the near future, cryptosystems that rely on the ECDL are no longer safe as the scalar multiple can be easily recovered using Shor's algorithm~\cite{shor1994algorithms}. 
Other quantum resilient schemes have been proposed. Furthermore, post-quantum cryptosystems such as Supersingular Isogeny Diffie-Hellman (SIDH)
 are slow techniques, and speeding up its elliptic curve computation is frequently mentioned as a significant goal.

The core operation for ECC is the scalar multiplication [k]P whose computation speed is seen as key to improving ciphers. For instant, in~\cite{eisentrager2003fast} Eisentrager et 
al proposed a method for computing the formula~$S = (2P + Q)$. Their improved procedure saves a field multiplication, when compared to the original algorithm. Later, Ciet et al~\cite{ciet2006trading} introduced a faster method for computing the same formula when a field inversion costs more than six field multiplications. 
Furthermore, they introduced an efficient method for computing point tripling. 
Mixed powers system of point doubling and tripling for computing the scalar multiplication was represented later by Dimitrov et al~\cite{dimitrov2005efficient}. In~\cite{mishra2007efficient} Mishra et al presented an efficient quintuple formula (5P) and introduced a mixed base algorithm with doubling and tripling. Further development was 
introduced by Longa and Miri~\cite{longa2008new} by computing an efficient method for 
tripling and quintupling mixed with differential addition. They proposed an efficient multibases non-adjacent representation (mbNAF) to reduce the cost. 
In~\cite{longa2008new} the same authors present further optimization in terms of cost for computing the form~$dP + Q$. 
They have succeeded in implementing the previous forms of mixed double and add algorithm by using a single inversion when applying a new precomputation scheme. 
More recently, Purohit and Rawat~\cite{purohit2011fast} used a multibase representation to propose an efficient scalar multiplication algorithm of doubling, tripling, and septupling, restricted on a non super-singular elliptic curve defined over the field~$F_{2^m}$. In addition, they have compared their work with other existing algorithms to achieve better representation in terms of cost. Therefore, speeding up the scalar multiplication computation in parallel with reducing the cost is a critical task. 
We present a new methodology to compute elliptic curve operations with more general forms of the type $mP+nQ$, 
where $m$ and $n$ are small integers, aiming for faster implementation with the lowest cost among currently known algorithms using only one inversion.

Among all applications based on EC, the highest benefit from our work concerns the post-quantum cryptosystem, Supersingular Isogeny Diffie-Helman (SIDH).
Its main weakness is the slow elliptic curve computation speed.
For other elliptic curve schemes, the computation speed-up also favors attacks, which can however be compensated by increasing the size of the key. Isogeny-based cryptography also utilizes points on an elliptic curve, but its security is instead based on the difficulty of computing isogenies between elliptic curves. An isogeny can be thought of as a unique algebraic mapping between two elliptic curves that satisfy the group law. An algorithm for computing isogenies on ordinary curves in sub-exponential time was presented by Childs et al~\cite{childs2014constructing}, rendering the use of cryptosystems based on isogenies on ordinary curves unsafe in the presence of quantum computers. However, there is no known algorithm for computing isogenies on supersingular curves in sub-exponential time. 

In~\cite{jao2011towards}, Jao and De Feo proposed a key exchange based on isogenies of supersingular elliptic curves. The proposed scheme resembles the standard Elliptic Curve Diffie-Hellman (ECDH), but goes a step further by computing isogenies over large degrees. 
In the scenario where Alice and Bob want to exchange a secret key over an insecure channel, 
they pick a smooth isogeny prime $p$ of the form $\l_{A}^{a}\l_{B}^{b}$ .~\textit{f} $\pm 1$ where $\l_{A}$ and $\l_{B}$ are small primes, $a$ and $b$ are positive integers, and \textit{f} is a small cofactor chosen to make the number prime. 
These define a supersingular elliptic curve, $\textit{E}_{0}(\textit{F}_{q})$ where $q = p^{2}$. 
Lastly, they choose four points on the curve that form the bases $\left \{ P_{A}, Q_{A}  \right \}$ and $\left \{ P_{B}, Q_{B}  \right \}$, which act as generators for $\textit{E}_{0}(\l_{A}^{a})$ and $\textit{E}_{0}(\l_{B}^{b})$, respectively. 
In a graph of supersingular isogenies where the vertices represent isomorphic curves and the edges represent l-degree isogenies, the infeasibility to discover a path that connects two particular vertices provides the security for this protocol. This led to the Supersingular Isogeny-based Diffie-Hellman key exchange protocol (SIDH)~\cite{jao2011towards}. As of today, the best-known algorithms against the SIDH protocol have an exponential time complexity for both classical and quantum attackers.

Although the SIDH public key size for achieving a 128-bit security level in the quantum setting was already reported as small as 564 bytes in~\cite{costello2016efficient}, this SIDH public key size was recently further reduced in~\cite{costello2017efficient} to just 330 bytes. 
However impressive, these key size credentials have to be contrasted against SIDH's relatively slow runtime performance. 
Indeed, the SIDH key exchange protocol has a latency in the order of milliseconds when implemented in high-end Intel processors. This timing is significantly higher than the one achieved by several other quantum-resistant cryptosystem proposals. Consequently, some recent works have focused on devising strategies to reduce the runtime cost of the SIDH protocol.

Reportedly Couveignes made the first suggestions towards the usage of isogenies for cryptographic purposes in a seminar held in 1997, later reported in~\cite{couveignes2006hard}. The first published work of a concrete isogeny-based cryptographic primitive was presented by Charles, Lauter and Goren in~\cite{cryptoeprint:2006:021,charles2009cryptographic}, where the authors introduced the hardness of path-finding in supersingular isogeny graphs and its application to the design of hash functions. It has since been used as an assumption for other cryptographic applications such as key-exchange and digital signature protocols. Stolbunov studied in~\cite{stolbunov2010constructing} the hardness of finding isogenies between two ordinary elliptic curves defined over a finite field $F_q$, with $q$ a prime power. The author proposed to use this setting as the underlying hard problem for a Diffie-Hellman-like key exchange protocol. Nevertheless, Childs, Jao, and Soukharev discovered in~\cite{childs2014constructing} a subexponential complexity quantum attack against Stolbunov’s scheme.

In 2011, Jao and De Feo proposed the problem of finding the isogeny map between two supersingular elliptic curves, a setting where the attack in~\cite{childs2014constructing} does not apply anymore. This proposal led to the Supersingular Isogeny-based Diffie-Hellman key exchange protocol (SIDH)~\cite{jao2011towards}. 
As of today, the best-known algorithms against the SIDH protocol have an exponential time complexity for both classical and quantum attackers.

On the other hand, some recent works have focused on devising strategies to reduce the runtime cost of the SIDH protocol. For example, Koziel et al. presented a parallel evaluation of isogenies implemented on an FPGA architecture~\cite{koziel2017post,koziel2016fast}, reporting important speedups for this protocol. These developments show the increasing research interest on developing techniques able to accelerate the SIDH protocol software and hardware implementations. More recently in~\cite{faz2018faster}, several algorithmic optimization targeting both elliptic-curve and field arithmetic operations have been presented in order to accelerate the runtime performance of SIDH by enhancing the calculation of the elliptic curve operation $P + k[Q]$.

The core operations for SIDH is computing the isogeny and its kernel. Basically, Velu’s formula is used to compute the isogeny and the $P + k[Q]$ formula is used to compute the kernel, where $P$ and $Q$ are points on the curve and $k$ is the secret key that is generated by both of them~\cite{jao2011towards}. This operation must be performed in both phases of SIDH. First, this happens in the key generation phase, where the point is known in advance. 
In this case, one can construct a look up table that contains all doubles of point $Q$ and reuse any of them when it’s needed. 
Second, in the key exchange phase, where the point $Q$ is variable, we will apply our mixed-base representation (up to 32) in order to speed up the calculations, minding that all mixed-base formulas was implemented with a single inversion.

\paragraph{Montgomery}
Here is another form of elliptic curve, different from the short Weierstrass form, which is called Montgomery curve. In 1987, the Montgomery curve was introduced by Peter L. Montgomery~\cite{montgomery1987speeding}. Such a curve is a set of all pairs (x,y)$\in$ $\Z_{q}$ which fulfill: 
\begin{align*}
E/Fq: By^2 = x^3 + Ax^2 + x
\end{align*}

such that A,B$\in$ $\Z_{q}$, and with $A^2\neq$4 and B is a non zero value, 

The point on an elliptic curve with affine coordinates can be represented in Montgomery form using projective coordinates P = (X : Z), where x = $X/Z$ and y = $Y/Z$ for Z$\neq$0.  

For $P=(X_1 : Z_1)$ and $Q=(X_2 : Z_2)$,
one can compute P + Q and point 2*P by using the following equations, where P $\neq$ Q:
\begin{align*}
X_3 = (Z_2-Z_1)((X_2-Z_2)(X_1+Z_1)+(X_2+Z_2)(X_1-Z_1))^2
\end{align*}
\begin{align*}
Z_3 = (X_2-X_1)((X_2-Z_2)(X_1+Z_1)-(X_2+Z_2)(X_1-Z_1))^2
\end{align*} 

In case of doubling where P = Q:
\begin{align*}
4X_1Z_1 = (X_1+Z_1)^2-(X_1-Z_1)^2
\end{align*}
\begin{align*}
X_3 = (X_1+Z_1)^2(X_1-Z_1)^2
\end{align*}
\begin{align*}
Z_3 = (4X_1Z_1)((X_1-Z_1)^2-((A+2)/4)(4X_1Z_1))
\end{align*}

\paragraph{Weierstrass Elliptic Curve}
This section represents the equations of the original work that we compare our algorithm with. We consider elliptic curve over $\Z_{p}$, where p~$>$~3. Such a curve, in the short Weierstrass form in the affine plane, is the set of all pairs (x,y)$\in$ $\Z_{p}$ which fulfill:

\begin{equation}
y^{2} \equiv x^{3} + a.x + b~~ (\rm{mod} ~~p) 
\end{equation}

For $P=(x_P,y_P)$ and $Q=(x_Q,y_Q)$,
one can compute $P + Q$ by using the following equations, where $\lambda$ is represented into two different forms \cite{paar2009understanding}. 

In case of addition where P $\neq$ Q:
\begin{equation}
\lambda = \left(\frac{y_Q-y_P}{x_Q-x_P}\right)~~ mod ~~p\label{eq:mc2.0}
\end{equation} 
\begin{equation}
x_R = \lambda^2-x_P-x_Q~~ mod ~~p \nonumber\label{eq:mc2.pq2}
\end{equation}

In case of computing $2*P$ (doubling of order one) where P has coordinates ($x_1$, $y_1$):
\begin{equation}
\lambda = \left(\frac{3x_1^2+a}{2y_1}\right)~~ mod ~~p\label{eq:mc2.1}
\end{equation} 
\begin{equation}
x_2 = \lambda^2-2x_1~~ mod ~~p \label{eq:mc2.2}
\end{equation}
\begin{equation}
y_2 = \lambda(x_1 - x_2)-y_1~~ mod ~~p \label{eq:mc2.3}
\end{equation}

Where $\lambda$ is the slope of the tangent through P, and $x_{2}$ and $y_{2}$, the affine coordinates after doubling P one time. While a two dimensional projective space can also be used for computations in Weierstrass form, here we focus on computations in the affine plane.

\section{Inverting Multiplications based on Curve Order}
Inverting multiplications can be very useful in case the number of 1 bits of the binary representation of an inversion of the scalar multiplication k is smaller than the number of 1 bits of the original one. Basically, instead of using the point P as a base, one traverses the group backward starting from the inversion of the base point $-P$. 

$$[k]P=[ord_E(P)-k](-P)=[\#E-k](-P)$$

If $\#E=61$ and $k=49$,
the number of 1-bit of the binary representation of $12_{10}=(1100)_2$ is less than the number of 1-bit of $49_{10}=(110001)_2$ by one, directing to a smaller number of differential additions. 

According to Hasse's theorem~\cite{paar2009understanding}:
\begin{align*}
p~+~1~-2\sqrt{p}~\leq~\#E~\leq~p~+1~+~2\sqrt{p}   
\end{align*} 

This is also known as Hasse's bound, which states that the number of points on elliptic curves is roughly in the range of the prime p.

\section{Fast $nP+mQ$}\label{sec:mn}
In this section we introduce step by step the proposed affine computations of the form $nP+mQ$.
We start with fast one-inversion formula for $2^nP$ (referred to as doubling of order $n$, or $n^{th}$ double), before addressing more complex equations.

\subsection{Fast $2^2P$}\label{sec:da}
By replacing $x_1,y_1$ and $x_{3}$, $y_{3}$, and $\lambda$ in  Equations~\ref{eq:mc2.1},~\ref{eq:mc2.2} and~\ref{eq:mc2.3}, we find the new slope and $(x_4,y_4)$ coordinates for the second double.

\begin{align*}
\lambda_{4}  = \frac{3x_{3}^2+a}{2y_3}~~ mod ~~p
\end{align*}
\begin{align*}
\lambda_{4}  = \frac{3(\lambda^{2}-2x_1)^2+a}{2(\lambda(x_{1}-(\lambda^{2}-2x_{1}))-y_{1})}~~ mod ~~p 
\end{align*}
\begin{align*}
\lambda_{4}  = \frac{3((\frac{3x_1^2+a}{2y_1})^{2}-2x_1)^2+a}{2(\lambda(x_{1}-(\lambda^{2}-2x_{1}))-y_{1})}~~ mod ~~p 
\end{align*}

\begin{equation}
\lambda_{4}  = \frac{3((\frac{3x_1^2+a}{2y_1})^{2}-2x_1)^2+a}{2x_1(\frac{3x_1^2+a}{2y_1})-2(\frac{3x_1^2+a}{2y_1})^3+4x_1(\frac{3x_1^2+a}{2y_1})-2y_1}~ mod ~p\label{eq:mc2.5} 
\end{equation}
~

In order to get rid of all inverses in Equation~\ref{eq:mc2.5} we multiply $\lambda_{4}$ by $\frac{(2y_1)^4}{(2y_1)^4}$ to eliminate all the denominators of the slope $\lambda$, then we get  
\begin{equation}
\lambda_{4}  = \frac{3((3x_1^2+a)^{2}-2x_1(2y_1)^2)^2+a(2y_1)^4}{2x_1(2y_1)^3(3x_1^2+a)-4y_1(3x_1^2+a)^3+4x_1(2y_1)^3(3x_1^2+a)-(2y_1)^5}~ mod ~p 
\end{equation}

The denominator is denoted U.
$$U=2x_1(2y_1)^3(3x_1^2+a)-4y_1(3x_1^2+a)^3+4x_1(2y_1)^3(3x_1^2+a)-(2y_1)^5$$

Rewritten as:
$$U=2y_1 (2x_1(2y_1)^2(3x_1^2+a)-2(3x_1^2+a)^3+4x_1(2y_1)^2(3x_1^2+a)-(2y_1)^4)$$

Further rewritten as:
\begin{equation}
U=2y_1 q \label{eq:mc2.U}
\end{equation}
where,
$$q=2x_1(2y_1)^2(3x_1^2+a)-2(3x_1^2+a)^3+4x_1(2y_1)^2(3x_1^2+a)-(2y_1)^4$$

Eliminating the inverses speeds up the calculation and increase the efficiency of the parallelization process between the slope and the coordinates equations. 

For simplicity, let's consider,
\begin{equation}
\lambda_{4}  = \frac{W}{U}~ mod ~p \label{eq:mc2.7}
\end{equation}

Then we substitute Equation~\ref{eq:mc2.7} in the $x_4$ and $y_4$ equations,
\begin{align*}
x_4 = \lambda_{4}^2-2x_3~~ mod ~~p
\end{align*}
\begin{align*}
x_4 = \lambda_{4}^2-2(\lambda^{2}-2x_1)~~ mod ~~p 
\end{align*}
\begin{equation}
x_4 = \lambda_{4}^2-2\lambda^{2}+4x_1~~ mod ~~p \label{eq:mc2.8}
\end{equation}
\begin{equation}
x_4 = \left(\frac{W}{U}\right)^2-2\left(\frac{3x_1^2+a}{2y_1}\right)^{2}+4x_1~~ mod ~~p \label{eq:mc2.9}
\end{equation}

Eliminating the inverses in Equation~\ref{eq:mc2.9} by multiplying with the value of $U^{2}$ where we remind from Equation~\ref{eq:mc2.U} that, 
\begin{equation*}
U = (2y_1)~q
\end{equation*}
Then we get,
\begin{align*}
U^{2}~x_4 = W^2-2q^2(3x_{1}^2+a)^2+4x_1U^2~~ mod ~~p
\end{align*}  
\begin{equation}
x_4 = \frac{W^2-2q^2(3x_{1}^2+a)^2+4x_1U^2}{U^2}~~ mod ~~p 
\end{equation}
 
Same steps will be applied in order to find and simplify $y_4$
\begin{align*}
y_4 = \lambda_{4}(x_3-x_4)-y_3~~ mod ~~p 
\end{align*}
\begin{align*}
y_4 = \lambda_{4}((\lambda^{2}-2x_1)-x_4)-(\lambda(x_1-(\lambda^{2}-2x_1))-y_1)~~ mod ~~p 
\end{align*}
\begin{align*}
y_4 = \lambda_{4}\lambda^{2}-2x_1\lambda_{4}-x_4\lambda_{4}-x_1\lambda+\lambda^{3}-2x_1\lambda+y_1~~ mod ~~p 
\end{align*}

Considering,
\begin{equation}
x_4 = \frac{N_x}{U^2}~~ mod ~~p\label{eq:mc2.12} 
\end{equation} 

\begin{align*}
y_4 =& \frac{W}{U}(\frac{3x_1^2+a}{2y_1})^2-2x_1\frac{W}{U}-x_4\frac{W}{U}\\
 &-x_1(\frac{3x_1^2+a}{2y_1})+(\frac{3x_1^2+a}{2y_1})^3-2x_1(\frac{3x_1^2+a}{2y_1})+y_1 
\end{align*}

Then we multiply $y_4$ by $U^3$

$U^3~y_4 = Wq^2(3x_1^2+a)^2-2x_1WU^2-N_xW-x_1U^2q(3x_1^2+a)+q^3(3x_1^2+a)^3-2x_1U^2q(3x_1^2+a)+y_1U^3$

\begin{equation}
y_4 = \frac{N_y}{U^3}~~ mod ~~p \label{eq:mc2.13}
\end{equation}

As it can be noted in Equations~\ref{eq:mc2.7},~\ref{eq:mc2.12}, and~\ref{eq:mc2.13}, the denominators of $x_4$ and $y_4$ are multiples of the $\lambda_{4}$ denominator. Therefore, one can implement the second double with only one inverse, unlike the original equations that involve two inverses in order to compute the second double of a point on a curve. Furthermore, finding the $\lambda_{4}$ value is not required anymore.

\subsection{Fast $2^3P$}\label{sec:8Q}
By applying the same steps that were followed in finding the second double, one can find the third double. First, employing the  $x_{4}$, $y_{4}$, and $\lambda_4$ in  Equations~\ref{eq:mc2.1},~\ref{eq:mc2.2}, and~\ref{eq:mc2.3} respectively. 

\begin{align*}
\lambda_{8}  = \frac{3x_{4}^2+a}{2y_4}~~ mod ~~p
\end{align*}
\begin{align*}
\lambda_{8}  = \frac{3(\lambda_{4}^2-2((\frac{3x_1^2+a}{2y_1})^{2}-2x_1))^2+a}{2(\lambda_{4}(((\frac{3x_1^2+a}{2y_1})^{2}-2x_1)-x_4)-((\frac{3x_1^2+a}{2y_1})(x_1-((\frac{3x_1^2+a}{2y_1})^{2}-2x_1))-y_1))}\\~~ mod ~~p 
\end{align*} 
~

For simplification we rewrite this as:
\begin{equation}
\lambda_{8}  = \frac{W_8}{U_8}~~ mod ~~p\label{eq:mc2.14}
\end{equation}

$W_8 = 3((\frac{W}{U})^2-2((\frac{3x_1^2+a}{2y_1})^{2}-2x_1))^2+a$
 
$U_8 = 2(\lambda_{4}(((\frac{3x_1^2+a}{2y_1})^{2}-2x_1)-x_4)-((\frac{3x_1^2+a}{2y_1})(x_1-((\frac{3x_1^2+a}{2y_1})^{2}-2x_1))-y_1))$

$U_8 = 2(\lambda_{4}(((\frac{3x_1^2+a}{2y_1})^{2}-2x_1)-(\lambda_{4}^2-2((\frac{3x_1^2+a}{2y_1})^{2}-2x_1)))-((\frac{3x_1^2+a}{2y_1})(x_1-((\frac{3x_1^2+a}{2y_1})^{2}-2x_1))-y_1))$

$U_8 = 2\lambda_{4}(\frac{3x_1^2+a}{2y_1})^{2}-4x_1\lambda_{4}-2\lambda_{4}^3+4\lambda_{4}(\frac{3x_1^2+a}{2y_1})^{2}-8x_1\lambda_{4}-2x_1(\frac{3x_1^2+a}{2y_1})+2(\frac{3x_1^2+a}{2y_1})^{3}-4x_1(\frac{3x_1^2+a}{2y_1})+2y_1$

$U_8 = 6(\frac{W}{U})(\frac{3x_1^2+a}{2y_1})^{2}-12x_1(\frac{W}{U})-6x_1(\frac{3x_1^2+a}{2y_1})-2(\frac{W}{U})^3+2(\frac{3x_1^2+a}{2y_1})^3+2y_1$

\begin{align*}
\lambda_{8} = \frac{3((\frac{W}{U})^2-2((\frac{3x_1^2+a}{2y_1})^{2}-2x_1))^2+a}{6(\frac{W}{U})(\frac{3x_1^2+a}{2y_1})^{2}-12x_1(\frac{W}{U})-6x_1(\frac{3x_1^2+a}{2y_1})-2(\frac{W}{U})^3+2(\frac{3x_1^2+a}{2y_1})^3+2y_1}~~ mod ~~p
\end{align*}

In order to eliminate all the inverses in the equation for $\lambda_8$, that are in $\lambda$ and $\lambda_4$, we multiply $\lambda_8$ by $\frac{U^4}{U^4}$, where,
\begin{equation}
U_8  = (2y_1)Uq_8~~ mod ~~p\label{eq:mc2.15}
\end{equation}

$W_8 = 3(W^2-2q^2(3x_1^2+a)^{2}+4x_1U^2)^2+aU^4~~ mod ~~p$

$U_8 = 6W(2y_1)q^3(3x_{1}^2+a)^2-12x_1WU^3-6x_1(2y_1)^3q^4(3x_{1}^2+a)-2W^3U+2(2y_1)q^4(3x_{1}^2+a)^3+(2y_1)U^4~~ mod ~~p$

Now, we reformulate the $U_8$ equation to maintain the form of the Equation~\ref{eq:mc2.15}. Thus, we have to multiply $\lambda_8$ (i.e., $\frac{W_8}{U_8}$), by $\frac{2y_1}{2y_1}$ then take out $(2y_1)U$ as a common factor of $U_8$. This way we can easily eliminate all inverses in $x_8$ and $y_8$.

$W_8 = (2y_1)(3(W^2-2q^2(3x_1^2+a)^{2}+4x_1U^2)^2+aU^4)~~ mod ~~p$ 

$U_8 = (2y_1)(6W(2y_1)q^3(3x_{1}^2+a)^2-12x_1WU^3-6x_1(2y_1)^3q^4(3x_{1}^2+a)-2W^3U+2(2y_1)q^4(3x_{1}^2+a)^3+(2y_1)U^4)~~ mod ~~p$

$U_8 = (2y_1)U(6Wq^2(3x_{1}^2+a)^2-12x_1WU^2-6x_1(2y_1)^2q^3(3x_{1}^2+a)-2W^3+2q^3(3x_{1}^2+a)^3+(2y_1)U^3)~~ mod ~~p$

where,

$q_8 = 6Wq^2(3x_{1}^2+a)^2-12x_1WU^2-6x_1(2y_1)^2q^3(3x_{1}^2+a)-2W^3+2q^3(3x_{1}^2+a)^3+(2y_1)U^3~~ mod ~~p$

To compute $x_8$ and $y_8$, substitute Equation~\ref{eq:mc2.14} with the new values of $W_8$ and $U_8$ in the $x_8$ and $y_8$ equations,

\begin{align*}
x_8 = \lambda_{8}^2-2x_4~~ mod ~~p
\end{align*}
\begin{align*}
x_8 = \lambda_{8}^2-2(\lambda_{4}^2-2(\lambda^{2}-2x_1))~~ mod ~~p
\end{align*}
\begin{align*}
x_8 = \lambda_{8}^2-2(\lambda_{4}^2-2\lambda^{2}+4x_1)~~ mod ~~p
\end{align*}
\begin{equation}
x_8 = \lambda_{8}^2-2\lambda_{4}^2+4\lambda^{2}-8x_1~~ mod ~~p\label{eq:mc2.16}
\end{equation}
\begin{align*}
x_8 = (\frac{W_8}{U_8})^2-2(\frac{W}{U})^2+4(\frac{3x_{1}^2+a}{2y_1})^2-8x_1~~ mod ~~p
\end{align*}
~

Multiply both sides by $U_{8}^2$ by considering the value of $U_8$ in~\ref{eq:mc2.15} to eliminate all inverses,
\begin{align*}
U_{8}^2~x_8 = W_{8}^2-2W^2(2y_1)^2q_{8}^2+4(3x_{1}^2+a)^2U^2q_{8}^2-8x_1U_{8}^2~~ mod ~~p
\end{align*} 

\begin{equation}
x_8 = \frac{W_{8}^2-2W^2(2y_1)^2q_{8}^2+4(3x_{1}^2+a)^2U^2q_{8}^2-8x_1U_{8}^2}{U_{8}^2}~~ mod ~~p
\end{equation}
~

Same steps will be applied in order to find and simplify $y_8$,
\begin{align*}
y_8 = \lambda_{8}(x_4-x_8)-y_4~~ mod ~~p 
\end{align*}

$y_8 = \lambda_{8}((\lambda_{4}^2-2(\lambda^{2}-2x_1))-x_8)-\lambda_{4}((\lambda^{2}-2x_1)-(\lambda_{4}^2-2(\lambda^{2}-2x_1)))+(\lambda(x_1-(\lambda^{2}-2x_1))-y_1)~~ mod ~~p$

$y_8 = \lambda_{8}(\lambda_{4}^2-2\lambda^{2}+4x_1-x_8)-(\lambda_{4}\lambda^{2}-2x_1\lambda_{4}-x_4\lambda_{4})+(\lambda x_1-\lambda^{3}+2x_1\lambda-y_1)~~ mod ~~p$

\begin{equation}
y_8 = \lambda_{8}\lambda_{4}^2-2\lambda_{8}\lambda^{2}+4x_1\lambda_{8}-x_8\lambda_{8}-\lambda_{4}\lambda^{2}+2x_1\lambda_{4}+x_4\lambda_{4}+\lambda x_1-\lambda^{3}+2x_1\lambda-y_1~~ mod ~~p\label{eq:mc2.18}
\end{equation}

Similarly to the decomposition $x_4 = \dfrac{N_x}{U^2}$, we consider $x_8 = \dfrac{N_{x8}}{U_{8}^2}$

$y_8 = \frac{W_8}{U_8}(\frac{W}{U})^2-2\frac{W_8}{U_8}(\frac{3x_{1}^2+a}{2y_1})^2+4x_1\frac{W_8}{U_8}-\frac{N_{x8}}{U_{8}^2}\frac{W_8}{U_8}-\frac{W}{U}(\frac{3x_{1}^2+a}{2y_1})^2+2x_1\frac{W}{U}+\frac{N_x}{U^2}\frac{W}{U}+x_1(\frac{3x_{1}^2+a}{2y_1})-(\frac{3x_{1}^2+a}{2y_1})^3+2x_1(\frac{3x_{1}^2+a}{2y_1})-y_1~~ mod ~~p$

Multiplying both sides by $U_{8}^3$,

$U_{8}^3~y_8 = W_8W^2(2y_1)^2q_{8}^2-2W_8(3x_{1}^2+a)^2U^2q_{8}^2+4x_1W_8U_{8}^2-N_{x8}W_8-W(3x_{1}^2+a)^2U_8Uq_{8}^2+2x_1WU_{8}^2(2y_1)q_8+N_xW(2y_1)^3q_{8}^3+x_1U_{8}^2Uq_8(3x_{1}^2+a)-U^3q_{8}^3(3x_{1}^2+a)^3+2x_1U_{8}^2Uq_8(3x_{1}^2+a)-y_1U_{8}^3~~ mod ~~p$ 

\begin{equation}
y_8 = \frac{N_{y8}}{U_{8}^3}~~ mod ~~p
\end{equation}
 
As it can be observed, this proves that one can compute the third double of a node on elliptic curve with only one inverse as we have done for the second double.

\subsection{Fast $2^4P$}\label{sec:16Q}
By expanding values $x_{8}$, $y_{8}$, and $\lambda_8$ in the $\lambda_{16}$ equation, we find the new slope for the fourth double. Then, we apply the same steps that were followed in the previous sections.
\begin{align*}
\lambda_{16}  = \frac{3x_{8}^2+a}{2y_8}~~ mod ~~p
\end{align*}

Considering,
\begin{equation}
\lambda_{16}  = \frac{W_{16}}{U_{16}}~~ mod ~~p
\end{equation}

where,
\begin{equation}
U_{16}  = (2y_1)U_8 U q_{16}~~ mod ~~p\label{eq:mc.212}
\end{equation}

Referring to the Equation~\ref{eq:mc2.16} for $x_8$ value,

$W_{16} = 3(\lambda_{8}^2-2\lambda_{4}^2+4\lambda^{2}-8x_1)^2+a~~ mod ~~p$

$W_{16} = 3((\frac{W_8}{U_8})^2-2(\frac{W}{U})^2+4(\frac{3x_{1}^2+a}{2y_1})^{2}-8x_1)^2+a~~ mod ~~p$

Referring to Equations~\ref{eq:mc2.8},~\ref{eq:mc2.16}, and~\ref{eq:mc2.18} for the values of $x_4$, $x_8$, and $y_8$ respectively, we get,

$U_{16}= 2(\lambda_{8}\lambda_{4}^2-2\lambda_{8}\lambda^{2}+4x_1\lambda_{8}-x_8\lambda_{8}-\lambda_{4}\lambda^{2}+2x_1\lambda_{4}+x_4\lambda_{4}+\lambda x_1-\lambda^{3}+2x_1\lambda-y_1)~~ mod ~~p$

$U_{16}= 2(\lambda_{8}\lambda_{4}^2-2\lambda_{8}\lambda^{2}+4x_1\lambda_{8}-(\lambda_{8}^2-2\lambda_{4}^2+4\lambda^{2}-8x_1)\lambda_{8}-\lambda_{4}\lambda^{2}+2x_1\lambda_{4}+(\lambda_{4}^2-2\lambda^{2}+4x_1)\lambda_{4}+\lambda x_1-\lambda^{3}+2x_1\lambda-y_1)~~ mod ~~p$

$U_{16}= 6\lambda_8\lambda_{4}^2-12\lambda^2\lambda_8+24x_1\lambda_8-2\lambda_{8}^3-6\lambda_4\lambda^2+12x_1\lambda_4+2\lambda_{4}^3+6x_1\lambda-2\lambda^3-2y_1~~ mod ~~p$

$U_{16}= 6(\frac{W_8}{U_8})(\frac{W}{U})^2-12(\frac{3x_{1}^2+a}{2y_1})^2(\frac{W_8}{U_8})+24x_1(\frac{W_8}{U_8})-2(\frac{W_8}{U_8})^3-6(\frac{W}{U})(\frac{3x_{1}^2+a}{2y_1})^2+12x_1(\frac{W}{U})+2(\frac{W}{U})^3+6x_1(\frac{3x_{1}^2+a}{2y_1})-2(\frac{3x_{1}^2+a}{2y_1})^3-2y_1~~ mod ~~p$

In order to eliminate all the inverses in the equation $\lambda_{16}$ that are part of $\lambda$, $\lambda_4$, and $\lambda_8$ we multiply $\lambda_8$ by $\frac{U_{8}^4}{U_{8}^4}$, while considering the value of $U_{8}$ in Equation~\ref{eq:mc2.15}.

$W_{16} = 3(W_8^2-2W^2(2y_1)^2q_{8}^2+4U^2q_{8}^2(3x_{1}^2+a)^{2}-8x_1U_{8}^2)^2+aU_{8}^4~~ mod ~~p$

$U_{16} = 6W_8W^2(2y_1)^2q_{8}^2U_8-12(3x_{1}^2+a)^2W_8q_{8}^2U^2U_{8}+24x_1W_8U_{8}^3-2W_{8}^3U_8-6W(3x_{1}^2+a)^2q_{8}^2UU_{8}^2+12x_1W(2y_1)q_8U_{8}^3+2W^3(2y_1)^3q_{8}^3U_8+6x_1(3x_{1}^2+a)q_8UU_{8}^3-2(3x_{1}^2+a)^3q_{8}^3U^3U_8-2y_1U_{8}^4~~ mod ~~p$

Now, we reformulate the equation of $U_{16}$ to maintain the form of Equation~\ref{eq:mc2.15}. Thus, we have to multiply $\lambda_{16}$ ($W_{16}$,$U_{16}$), by $\frac{(2y_1)U}{(2y_1)U}$ then take out $(2y_1)UU_8$ as a common factor of $U_{16}$. This way we can eliminate all inverses from the equations for computing $x_{16}$ and $y_{16}$.

$W_{16} = (2y_1)U(3(W_8^2-2W^2(2y_1)^2q_{8}^2+4U^2q_{8}^2(3x_{1}^2+a)^{2}-8x_1U_{8}^2)^2+aU_{8}^4)~~ mod ~~p$

$U_{16} = (2y_1)U(6W_8W^2(2y_1)^2q_{8}^2U_8-12(3x_{1}^2+a)^2W_8q_{8}^2U^2U_{8}+24x_1W_8U_{8}^3-2W_{8}^3U_8-6W(3x_{1}^2+a)^2q_{8}^2UU_{8}^2+12x_1W(2y_1)q_8U_{8}^3+2W^3(2y_1)^3q_{8}^3U_8+6x_1(3x_{1}^2+a)q_8UU_{8}^3-2(3x_{1}^2+a)^3q_{8}^3U^3U_8-2y_1U_{8}^4)~~ mod ~~p$

$U_{16} = (2y_1)UU_8(6W_8W^2(2y_1)^2q_{8}^2-12(3x_{1}^2+a)^2W_8q_{8}^2U^2+24x_1W_8U_{8}^2-2W_{8}^3-6W(3x_{1}^2+a)^2q_{8}^2UU_{8}+12x_1W(2y_1)q_8U_{8}^2+2W^3(2y_1)^3q_{8}^3+6x_1(3x_{1}^2+a)q_8UU_{8}^2-2(3x_{1}^2+a)^3q_{8}^3U^3-2y_1U_{8}^3)~~ mod ~~p$

Thus one finds $\lambda_{16}$ with a single inverse, unlike the original method that requires 4 inverses in order to compute the fourth double slope. Likewise, we prove in the following equations that we can find $x_{16}$ and $y_{16}$ with a multiplier denominator of $\lambda_{16}$. As a result, there is no need to calculate their inverses as we have proven before. 
\begin{align*}
x_{16} = \lambda_{16}^2 - 2x_8~~ mod ~~p
\end{align*}
\begin{align*}
x_{16} = \lambda_{16}^2 - 2(\lambda_{8}^2-2(\lambda_{4}^2-2(\lambda^{2}-2x_1)))~~ mod ~~p
\end{align*}
\begin{align*}
x_{16} = \lambda_{16}^2 - 2(\lambda_{8}^2-2(\lambda_{4}^2-2\lambda^{2}+4x_1))~~ mod ~~p
\end{align*}
\begin{align*}
x_{16} = \lambda_{16}^2 - 2(\lambda_{8}^2-2\lambda_{4}^2+4\lambda^{2}-8x_1)~~ mod ~~p
\end{align*}
\begin{align*}
x_{16} = \lambda_{16}^2 - 2\lambda_{8}^2+4\lambda_{4}^2-8\lambda^{2}+16x_1~~ mod ~~p
\end{align*}
\begin{align*}
x_{16} = (\frac{W_{16}}{U_{16}})^2-2(\frac{W_{8}}{U_{8}})^2+4(\frac{W}{U})^2-8(\frac{3x_{1}^2+a}{2y_1})^{2}+16x_1~~ mod ~~p
\end{align*}

In order to eliminate the inverses in the equation of $x_{16}$, we multiply both sides by $U_{16}^2$, using the value of $U_{16}$ in Equation~\ref{eq:mc.212},\\[0.5cm]
$U_{16}^2~x_{16} = W_{16}^2-2W_{8}^2(2y_1)^2U^2q_{16}^2+$\\ 
\begin{align*}
4W^2(2y_1)^2U_{8}^2q_{16}^2-8(3x_{1}^2+a)^{2}U_{8}^2U^2q_{16}^2+16x_1U_{16}^2~~ mod ~~p
\end{align*}
$x_{16} =$
\begin{align*}
\frac{W_{16}^2-2W_{8}^2(2y_1)^2U^2q_{16}^2+4W^2(2y_1)^2U_{8}^2q_{16}^2-8(3x_{1}^2+a)^{2}U_{8}^2U^2q_{16}^2+16x_1U_{16}^2}{U_{16}^2}
\end{align*}
\begin{equation}
~~~~~~~~~~~~~~~~~~~~~~~~~~~~~~~~~~~~~~~~~~~~~~~~~~~~~~~~~~~~~~~~ mod ~~p
\end{equation}

Likewise, finding $y_{16}$,
\begin{align*}
y_{16} = \lambda_{16}(x_8-x_{16})-y_8~~ mod ~~p
\end{align*}

Based on the decomposition of $x_8$ as $\frac{N_{x8}}{U_{8}^2}$, given that $y_8 = \frac{N_{y8}}{U_{8}^3}$, and rewriting $x_{16}$ as $\frac{N_{x16}}{U_{16}^2}$, we substitute with these values in the previous equation to find $y_{16}$. 
\begin{align*}
y_{16} = \frac{W_{16}}{U_{16}}(\frac{N_{x8}}{U_{8}^2}-\frac{N_{x16}}{U_{16}^2})-\frac{N_{y8}}{U_{8}^3}~~ mod ~~p
\end{align*}
 
Multiplying both sides by $U_{16}^3$, to eliminate all inverses as we have done previously with considering the value of $U_{16}$ in Equation~\ref{eq:mc.212},
\begin{align*}
U_{16}^3~y_{16} = W_{16}N_{x8}(2y_1)^2U^2q_{16}^2-W_{16}N_{x16}-N_{y8}(2y_1)^3U^3q_{16}^3~~ mod ~~p
\end{align*}
\begin{equation}
y_{16} = \frac{W_{16}N_{x8}(2y_1)^2U^2q_{16}^2-W_{16}N_{x16}-N_{y8}(2y_1)^3U^3q_{16}^3}{U_{16}^3}~~ mod ~~p
\end{equation}

\subsection{Fast 3P (Point Tripling)}\label{sec:3d}
As it is important to calculate the binary multiplicative $2^{n}$ for  points Q to compute a large degree isogeny, we enhance the algorithm by finding the intermediate steps like 3Q, 5Q, and 7Q etc.

In~\cite{rao2016three} Subramanya Rao have worked on Montgomery curves and found an efficient technique to find point tripling. Simply, we will optimize an application of a single double to some point Q then perform a point addition. This technique could be applied to all intermediate steps as follows. 

Substitute the Equation~\ref{eq:mc2.1} of the value of $\lambda$ in Equations~\ref{eq:mc2.2} and~\ref{eq:mc2.3}.
\begin{align*}
x_{2}=(\frac{3x_{1}^2+a}{2y_1})^2-2x_1~~ mod ~~p
\end{align*}
\begin{align*}
y_{2}=(\frac{3x_{1}^2+a}{2y_1})(x_1-x_2)-y_1~~ mod ~~p 
\end{align*}

Then, we substitute the value of $x_3$ and $y_3$ in the slope of point addition as a value of 2Q.
\begin{equation}
\lambda'_3 = \frac{y'_2-y'_1}{x'_2-x'_1}~~ mod ~~p
\nonumber
\end{equation}
Getting
\begin{align*}
\lambda'_3 = \frac{((\frac{3x_{1}^2+a}{2y_1})(x_1-x_3)-y_1)-y_1}{((\frac{3x_{1}^2+a}{2y_1})^2-2x_1)-x_1}~~ mod ~~p
\end{align*}
\begin{align*}
\lambda'_3 = \frac{(\frac{3x_{1}^2+a}{2y_1})(x_1-(\frac{3x_{1}^2+a}{2y_1})^2+2x_1)-2y_1}{((\frac{3x_{1}^2+a}{2y_1})^2-2x_1)-x_1}~~ mod ~~p
\end{align*}
\begin{align*}
\lambda'_3 = \frac{(\frac{3x_{1}^2+a}{2y_1})(3x_1-(\frac{3x_{1}^2+a}{2y_1})^2)-2y_1}{(\frac{3x_{1}^2+a}{2y_1})^2-3x_1}~~ mod ~~p
\end{align*}

Multiplying $\lambda'_3$ with $(\frac{2y_1}{2y_1})^3$ to eliminate all inverses and to be able to take out $2y_1$ as a common factor from the denominator to reduce the $x_2$ fraction which is the x-coordinate of the point doubling.
\begin{equation}
\lambda'_3 = \frac{(3x_{1}^2+a)(3x_1(2y_1)^2-(3x_{1}^2+a)^2)-(2y_1)^4}{(2y_1)((3x_{1}^2+a)^2-3x_1(2y_1)^2)}~~ mod ~~p
\end{equation}
Rewriting,
\begin{equation}
\lambda'_3 = \frac{W_3}{U_3}~~ mod ~~p\label{eq:mc2.lambda3}
\end{equation}
Where,
\begin{equation}
U_3 = (2y_1)q_3~~ mod ~~p
\end{equation}
Now we substitute with the value of the new slope in Equation~\ref{eq:mc2.lambda3}
to compute 3Q.
\begin{align*}
x_3 = {\lambda'}_3^2 - x_1 - x_2~~ mod ~~p
\end{align*}
\begin{align*}
x_3 = (\frac{W_3}{U_3})^2-x_1-(\frac{3x_{1}^2+a}{2y_1})^2+2x_1 ~~ mod ~~p
\end{align*}
Multiplying the equation with $U_{3}^2$
\begin{align*}
U_{3}^2~x_3 = W_3^2+x_1U_{3}^2-q_{3}^2(3x_{1}^2+a)^2~~ mod ~~p
\end{align*}
\begin{equation}
x_3 = \frac{W_3^2+x_1U_{3}^2-q_{3}^2(3x_{1}^2+a)^2}{U_{3}^2}~~ mod ~~p\label{eq:mc2.28}
\end{equation}
Considering, 
\begin{equation}
x_3 = \frac{N_{x3}}{U_{3}^2}~~ mod ~~p
\end{equation}
Now we find $y_3$,
\begin{align*}
y_3 = \lambda'_3 (x_1 - x_3) - y_1~~ mod ~~p
\end{align*}
\begin{align*}
y_3 = \frac{W_3}{U_3} (x_1 - \frac{N_{x3}}{U_{3}^2}) - y_1~~ mod ~~p
\end{align*}
Multiplying with $U_{3}^3$ to eliminate the inverses, 
\begin{align*}
U_{3}^3~y_3 = W_3 (x_1U_{3}^2 - N_{x3}) - y_1U_{3}^3~~ mod ~~p
\end{align*}
\begin{equation}
y_3 = \frac{W_3 (x_1U_{3}^2 - N_{x3}) - y_1U_{3}^3}{U_{3}^3}~~ mod ~~p
\end{equation}

\subsection{Fast $2^nQ+P$}\label{sec:plusp}
As we have mentioned earlier, the complexity of the SIDH cryptosystem relies on computing isogenies between points on the elliptic curve. Thus, we have performed a further optimization in term of the kernel Equation~$P + [k]Q$. As we have succeeded to perform an advanced exponent of a point on a curve with a single inverse, it would have been needed to compute an extra inverse for a differential point addition. Therefore, in this section we introduce an optimization for mixing our advanced doubling equations with the addition and perform it with a single inverse. 

In the point tripling section we computed the 3P as an intermediate step. Here, we provide general equations that can be applied to any of our equations 4Q, 8Q, and 16Q and their extensions. 

The following equations have some variables like $N_x$, $N_y$, and U that have to be replaced with the variables that related to each double. We have represented here the equation P + 4Q. 

We substitute the value of x and y coordinates of the second double of the point Q in equations~\ref{eq:mc2.12} and~\ref{eq:mc2.13} respectively in the addition slope equation in~\ref{eq:mc2.0}.
\begin{align*}
\lambda_5 = \frac{\frac{N_y}{U^3}-y_1}{\frac{N_x}{U^2}-x_1}~~ mod ~~p
\end{align*}
    
Multiplying with $U^3$ to eliminate the inverses,
\begin{equation}
\lambda_5 = \frac{N_y-y_1U^3}{N_xU-x_1U^3}~~ mod ~~p
\end{equation}

Substitute $\lambda$ in the equations for $x_5$ and $y_5$,
\begin{align*}
x_{5}=(\frac{N_y-y_1U^3}{N_xU-x_1U^3})^2-x_1-\frac{N_x}{U^2}~~ mod ~~p
\end{align*}

Take out a common factor U of the $\lambda$ denominator and multiply the equation with $U^2(N_x-x_1U^2)^2$,
\begin{align*}
U^2(N_x-x_1U^2)^2~x_{3}=(N_y-y_1U^3)^2-x_1U^2(N_x-x_1U^2)^2-N_x(N_x-x_1U^2)^2~~ mod ~~p
\end{align*} 
\begin{equation}
x_{3}=\frac{(N_y-y_1U^3)^2-x_1U^2(N_x-x_1U^2)^2-N_x(N_x-x_1U^2)^2}{U^2(N_x-x_1U^2)^2}~~ mod ~~p\label{eq:xplusp}
\end{equation}

Now we find $y_3$,
\begin{align*}
y_5 = \frac{N_y-y_1U^3}{U(N_x-x_1U^2)}x_1-\frac{N_y-y_1U^3}{U(N_x-x_1U^2)}x_3- y_1~~ mod ~~p
\end{align*}

Multiplying with $U^2(N_x-x_1U^2)^2$ one eliminates all inverses and adjusts the $y_3$ denominator to be matched to $x_5$ for simplification,
\begin{align*}
U^2(N_x-x_1U^2)^2~y_3 = U(N_x-x_1U^2)((N_y-y_1U^3)x_1-(N_y-y_1U^3)x_3- U(N_x-x_1U^2)y_1)
\end{align*}
\begin{equation}
y_5 = \frac{U(N_x-x_1U^2)((N_y-y_1U^3)x_1-(N_y-y_1U^3)x_3- U(N_x-x_1U^2)y_1)}{U^2(N_x-x_1U^2)^2}~~ mod ~~p
\end{equation}

\subsection{Fast $2^{n}$P + 2Q}\label{sec:plus2p}
In Section~\ref{sec:3d} we have illustrate the importance of computing the intermediate equations for the overall speed-up of our algorithms. The $2^{n}P+2Q$ form, where~$1 < n < 5$, will be efficient in computing the points 6Q, 10Q, and 18Q and since the subtraction in elliptic curve could be represented by adding the inverse of a point, 14Q can be performed efficiently by using this algorithm as well. In this section, we will exemplify implementing the form of $2^{2}P+2Q$=(6Q), where $P = Q$.

We substitute the value of x and y coordinates of the 4Q algorithm of the point Q in equations~\ref{eq:mc2.12} and~\ref{eq:mc2.13} respectively and the 2Q algorithm coordinates in Equations~\ref{eq:mc2.2} and~\ref{eq:mc2.3} as well in the addition slope Equation~\ref{eq:mc2.0}. 
\begin{align*}
\lambda_6 = \frac{\frac{N_{y^n}}{U^3}-(\frac{3x_{1}^2+a}{2y_1})(x_1-(\frac{3x_{1}^2+a}{2y_1})^2+2x_1)+y_1}{\frac{N_x}{U^2}-(\frac{3x_{1}^2+a}{2y_1})^2+2x_1}~~ mod ~~p
\end{align*}
    
Multiplying with $U^3$ to eliminate the inverses by considering the value of U in Equation~\ref{eq:mc2.U},
\begin{align*}
\lambda_6 = \frac{N_{y^n}-q^3(3x_{1}^2+a)(x_1(2y_1)^2-(3x_{1}^2+a)^2+2x_1(2y_1)^2)+y_1U^3}{N_{x^n}U-2y_1q^3(3x_{1}^2+a)^2+2x_1(2y_1)(2y_1)^2q3}~~ mod ~~p
\end{align*}
\begin{align*}
\lambda_6 = \frac{N_{y^n}-q^3(3x_{1}^2+a)(x_1(2y_1)^2-(3x_{1}^2+a)^2+2x_1(2y_1)^2)+y_1U^3}{N_{x^n}U-2y_1q^3(3x_{1}^2+a)^2+2x_1(2y_1)(2y_1)^2q3}~~ mod ~~p
\end{align*}
Since $U = 2y_1q$, we take a common factor U from the denominator then multiply both the numerator and denominator with $2y_1$, in order to be able to eliminate all inverses in the coordinates equations,
\begin{align*}
\lambda_6 = \frac{2y_1(N_y-q^3(3x_{1}^2+a)(x_1(2y_1)^2-(3x_{1}^2+a)^2+2x_1(2y_1)^2)+y_1U^3)}{2y_1U(N_{x^n}-q^2((3x_{1}^2+a)^2+2x_1(2y_1)^2))}~~ mod ~~p
\end{align*}
Considering,
\begin{equation}
\lambda_6 = \frac{W_6}{U_6}~~ mod ~~p
\end{equation}
Where,
\begin{equation}
U_6 = (2y_1)Uq_6~~ mod ~~p
\end{equation}
Now we substitute with the value of new slope in x and y coordinate equations to compute 6Q,
\begin{align*}
x_6 = \lambda_{6}^2 - x_3 - x_4~~ mod ~~p
\end{align*}
\begin{align*}
x_6 = (\frac{W_6}{U_6})^2-(\frac{3x_{1}^2+a}{2y_1})^2+2x_1-\frac{N_{x^n}}{U^2} ~~ mod ~~p
\end{align*}
Multiplying the equation with $U_{6}^2$
\begin{align*}
U_{6}^2~x_6 = W_{6}^2-U^2q_{6}^2(3x_{1}^2+a)^2+2x_1U_{6}^2-(2y_1)^2q_{6}^2N_{x^n}~~ mod ~~p
\end{align*}
\begin{equation}
x_6= \frac{W_{6}^2-U^2q_{6}^2(3x_{1}^2+a)^2+2x_1U_{6}^2-(2y_1)^2q_{6}^2N_{x^n}}{U_{6}^2}~~ mod ~~p
\end{equation}
Now we find $y_6$,
\begin{align*}
y_6 = \lambda_6 (x_3 - x_6) - y_3~~ mod ~~p
\end{align*}
Substituting the value of $y_3$, $x_3$ and $x_6$ by considering,
\begin{align*}
x_6 = \frac{N_{x6}}{U_{6}^2}
\end{align*}
\begin{align*}
y_6 = \frac{W_6}{U_6}((\frac{3x_{1}^2+a}{2y_1})^2-2x_1-\frac{N_{x6}}{U_{6}^2})-(\frac{3x_{1}^2+a}{2y_1})(x_1-(\frac{3x_{1}^2+a}{2y_1})^2+2x_1)+y_1 ~~ mod ~~p
\end{align*}
\begin{align*}
y_6 = \frac{W_6}{U_6}((\frac{3x_{1}^2+a}{2y_1})^2-2x_1-\frac{N_{x6}}{U_{6}^2})-(\frac{3x_{1}^2+a}{2y_1})(3x_1-(\frac{3x_{1}^2+a}{2y_1})^2)+y_1 ~~ mod ~~p
\end{align*}
Multiplying the equation with $U_{6}^3$ to eliminate the inverses, 

$U_{6}^3~y_6=W_6(U^2q_{6}^2(3x_{1}^2+a)^2-2x_1U_{6}^2-N_{x6})-Uq_6(3x_{1}^2+a)(3x_1U_{6}^2-U^2q_{6}^2(3x_{1}^2+a)^2)+y_1U_{6}^3~~ mod ~~p$
\begin{equation}
y_3 = \frac{N_{y6}}{U_{6}^3}~~ mod ~~p
\end{equation}

Note: In case of n = 3 or 4, $N_{x^n}$, $N_{y^n}$ and $U_n$ will be replaced with the values that related to each algorithm. For the variable q, it will be replaced with $Uq_8$ and $UU_8q_{16}$ in case of computing $8P+2Q$ and $16P+2Q$ respectively.

\subsection{$2^{n}$P + mQ}\label{sec:plus3p}
We have succeeded in representing this algorithms for some integers n and m, in one implementation with some replaceable variables, $N_{y^n}$, $N_{x^n}$, and $U_n$ based on the n value, where~($1 < n < 5$) and ($2 < m < 17$). As we have described in the previous sections,  each algorithm computes these variables differently. Thus, generalizing the equations benefits hardware optimization, flexibility and allows parallelization to be applied intensively. In this section, we illustrate the case of $2^3P+3Q$ = (11Q), where $P = Q$.

We substitute the value of the coordinates x and y of the 8P, 3P algorithms of the point P from Sections~\ref{sec:8Q}, and~\ref{sec:3d} respectively in the addition slope Equation~\ref{eq:mc2.0}. 

\begin{align*}
\lambda_{11} = \frac{\frac{N_{y8}}{U_{8}^3}-\frac{N_{y3}}{U_{3}^3}}{\frac{N_{x8}}{U_{8}^2}-\frac{N_{x3}}{U_{3}^2}}~~ mod ~~p
\end{align*}
    
Multiplying the numerator and denominator with $U_{8}^3U_{3}^3$ to eliminate the inverses,
\begin{align*}
\lambda_{11} = \frac{N_{y8}U_{3}^3-N_{y3}U_{8}^3}{N_{x8}U_{3}^3U_8-N_{x3}U_{8}^3U_3}~~ mod ~~p
\end{align*}
Take out a common factor $U_8U_3$ of the denominator,  
\begin{align*}
\lambda_{11} = \frac{N_{y8}U_{3}^3-N_{y3}U_{8}^3}{U_3U_8(N_{x8}U_{3}^2-N_{x3}U_{8}^2)}~~ mod ~~p
\end{align*}
Considering, 
\begin{equation}
\lambda_{11} = \frac{W_{11}}{U_{11}}~~ mod ~~p
\end{equation}
where,
\begin{equation}
U_{11} = U_3U_8q_{11}~~ mod ~~p
\end{equation}
Now we substitute with the value of new slope in x and y coordinate equations to compute 11Q,
\begin{align*}
x_{11} = \lambda_{11}^2 - x_3 - x_8~~ mod ~~p
\end{align*}
\begin{align*}
x_{11} = (\frac{W_{11}}{U_{11}})^2-\frac{N_{x3}}{U_{3}^2}-\frac{N_{x8}}{U_{8}^2} ~~ mod ~~p
\end{align*}
Multiplying the equation with $U_{11}^2$,
\begin{align*}
U_{11}^2~x_{11} = W_{11}^2-N_{x3}U_{8}^2q_{11}^2-N_{x8}U_{3}^2q_{11}^2~~ mod ~~p
\end{align*}
\begin{equation}
x_{11} = \frac{W_{11}^2-N_{x3}U_{8}^2q_{11}^2-N_{x8}U_{3}^2q_{11}^2}{U_{11}^2}~~ mod ~~p\label{eq:x11}
\end{equation}
where,
\begin{equation}
x_{11} = \frac{N_{x11}}{U_{11}^2}~~ mod ~~p
\end{equation}
Now we find $y_{11}$,
\begin{align*}
y_{11} = \lambda_{11} (x_3 - x_{11}) - y_3~~ mod ~~p
\end{align*}
Substitute the value of $y_3$, $x_3$ and $x_{11}$,
\begin{align*}
y_{11} = \frac{W_{11}}{U_{11}} \left(\frac{N_{x3}}{U_{3}^2} - \frac{N_{x11}}{U_{11}^2}\right) - \frac{N_{y3}}{U_{3}^3}~~ mod ~~p
\end{align*}
Multiplying the equation with $U_{11}^3$,   
\begin{align*}
U_{11}^3~y_{11} = W_{11} (N_{x3}U_{8}^2q_{11}^2 - N_{x11}) - N_{y3}U_{8}^3q_{11}^3~~ mod ~~p
\end{align*}
\begin{equation}
y_{11} = \frac{W_{11} (N_{x3}U_{8}^2q_{11}^2 - N_{x11}) - N_{y3}U_{8}^3q_{11}^3}{U_{11}^3}~~ mod ~~p
\end{equation}

\subsection{Generalizing $2^nP+Q$ and $2^nP+mQ$ Forms}\label{sec:3.5.7}
We notice that by reforming the equations in Section~\ref{sec:plusp}, we can generalize the two forms $2^nP+Q$ and $2^nP+mQ$.

As done in Section~\ref{sec:plusp}, we substitute the value of x and y coordinates of the second double of the point Q in Equations~\ref{eq:mc2.12} and~\ref{eq:mc2.13} respectively in the addition slope Equation~\ref{eq:mc2.0}.
\begin{align*}
\lambda = \frac{\frac{N_y}{U^3}-y_1}{\frac{N_x}{U^2}-x_1}~~ mod ~~p
\end{align*}
    
Multiplying with $U^3$ to eliminate the inverses, yields
\begin{equation}
\lambda = \frac{N_y-y_1U^3}{N_xU-x_1U^3}~~ mod ~~p
\end{equation} 

Considering,
\begin{equation}
\lambda_g = \frac{W_g}{U_g}~~ mod ~~p
\end{equation}

Substituting $\lambda$ in the equations for $x_g$ and $y_g$,
\begin{align*}
x_{g}=(\frac{W_g}{U_g})^2-x_1-\frac{N_x}{U^2}~~ mod ~~p
\end{align*}

Multiplying this equation with $U_{g}^2$ where,
\begin{equation}
U_g = Uq_g~~ mod ~~p\label{eq:Ug}
\end{equation}
\begin{align*}
U_{g}^2~x_{g}=W_g^2-x_1U_{g}^2-N_xq_{g}^2~~ mod ~~p
\end{align*}
\begin{equation}
x_{g}=\frac{W_g^2-x_1U_{g}^2-N_xq_{g}^2}{U_{g}^2}~~ mod ~~p
\end{equation}

Considering,
\begin{equation}
x_{g}=\frac{N_{xg}}{U_{g}^2}~~ mod ~~p\label{eq:xg}
\end{equation}

Now we find $y_g$,
\begin{align*}
y_g = \frac{W_g}{U_g}(x_1-\frac{N_{xg}}{U_{g}^2}) - y_1~~ mod ~~p
\end{align*}

Multiplying the equation with $U_{g}^2$ in order to eliminate all inverses, 
\begin{align*}
U_{g}^3~y_g = W_g(x_1U_{g}^2-N_{xg}) - y_1U_{g}^3~~ mod ~~p
\end{align*}

\begin{equation}
y_g = \frac{W_g(x_1U_{g}^2-N_{xg}) - y_1U_{g}^3}{U_{g}^3}~~ mod ~~p
\end{equation}

The previous equations have variables $N_{xg}$, $N_{yg}$, $U_g$'s and $W_g$'s that have to be replaced with the variables that related to the corresponding double. Additionally, as we have seen in Equations~\ref{eq:xplusp} and~\ref{eq:x11} of coordinate x, the terms $x_1U_{g}^2$, and $N_xq_{g}^2$ have to be replaced with~$N{xi}U_{j}^2q_{i+j}^2$, and $N_{xj}U_{i}^2q_{i+j}^2$ respectively in case we use the $2^nP+mQ$ form. As well as in the equation of coordinate y, we replace the terms $x_1U_{g}^2$, and $y_1U_{g}^3$ respectively with~$N_{xi}U_{j}^2q_{i+j}^2$, and $N_{yi}U_{j}^3q_{i+j}^3$. Knowingly, we have succeeded to generalize these two forms with replacing some variables and terms, leading to a potential optimization in terms of hardware.    

\subsection{Another Implementation of 6Q}\label{sec:6q}
In Section~\ref{sec:plus2p}, we have illustrate that we can compute 6Q through the form of $2^2P + 2Q$. In this section, we provide a different algorithm for computing 6Q by doubling the point 3Q which has been discussed in Section~\ref{sec:3d}. Diversity in implementations provides us with a greater chance of comparison in terms of hardware versus speed.

Since it's a doubling operation, we substitute the value of x and y coordinates that has been successfully computed in Section~\ref{sec:3d} in the slope Equation~\ref{eq:mc2.1},
\begin{align*}
\lambda_{6} = \left(\frac{3x_3^2+a}{2y_3}\right)~~ mod ~~p 
\end{align*}
\begin{align*}
\lambda_{6} = \frac{3(\frac{W_3^2+x_1U_{3}^2-q_{3}^2(3x_{1}^2+a)^2}{U_{3}^2})^2+a}{2(\frac{U_3W_3 (x_1 - x_3) - y_1U_{3}^2}{U_{3}^2})}~~ mod ~~p 
\end{align*}
Substituting the $x_3$ value in the $y_3$,
\begin{align*}
\lambda_{6} = \frac{3(\frac{W_3^2+x_1U_{3}^2-q_{3}^2(3x_{1}^2+a)^2}{U_{3}^2})^2+a}{2(\frac{U_3W_3x_1 - W_3(\frac{W_3^2+x_1U_{3}^2-q_{3}^2(3x_{1}^2+a)^2}{U_{3}}) - y_1U_{3}^2}{U_{3}^2})}~~ mod ~~p 
\end{align*}
Multiplying with $\frac{U_{3}^4}{U_{3}^4}$,
\begin{align*}
\lambda_{6} = \frac{3(W_3^2+x_1U_{3}^2-q_{3}^2(3x_{1}^2+a)^2)^2+aU_{3}^4}{2U_{3}^3W_3x_1 - 2U_3W_3(W_{3}^2+x_1U_{3}^2-q_{3}^2(3x_{1}^2+a)^2) - 2y_1U_{3}^4}~~ mod ~~p 
\end{align*}
Take out a common factor $U_3$ of the denominator in order to eliminate the x and y inverses in the new coordinates equations,
\begin{align*}
\lambda_{6} = \frac{3(W_3^2+x_1U_{3}^2-q_{3}^2(3x_{1}^2+a)^2)^2+aU_{3}^4}{U_{3}(2U_{3}^2W_3x_1 - 2W_3(W_{3}^2+x_1U_{3}^2-q_{3}^2(3x_{1}^2+a)^2) - 2y_1U_{3}^3)}~~ mod ~~p 
\end{align*}
Considering,
\begin{equation}
\lambda_{6} = \frac{W_6}{U_6}~~ mod ~~p\label{eq:mc2.43}
\end{equation}
where,
\begin{equation}
U_{6} = U_3q_6~~ mod ~~p\label{eq:mc2.44}
\end{equation}
Now we substitute the value of $\lambda_{6}$ in Equation~\ref{eq:mc2.43} with considering the value of $U_{6}$ in Equation~\ref{eq:mc2.44} when we eliminate the inverses,
\begin{align*}
x_{6} = \lambda_{6}^2 - 2x_3~~ mod ~~p
\end{align*}
\begin{align*}
x_{6} = \left(\frac{W_6}{U_6}\right)^2 - 2\left(\frac{W_3^2+x_1U_{3}^2-q_{3}^2(3x_{1}^2+a)^2}{U_{3}^2}\right)~~ mod ~~p
\end{align*}
Multiplying the equation with $U_{6}^2$,
\begin{align*}
U_{6}^2~x_{6} = W_6^2 - 2q_{6}^2(W_3^2+x_1U_{3}^2-q_{3}^2(3x_{1}^2+a)^2)~~ mod ~~p
\end{align*}
\begin{equation}
x_{6} = \frac{W_6^2 - 2q_{6}^2(W_3^2+x_1U_{3}^2-q_{3}^2(3x_{1}^2+a)^2)}{U_{6}^2}~~ mod ~~p\label{eq:mc2.45}
\end{equation}
Now we find $y_6$,
\begin{align*}
y_{6} = \lambda_{6} (x_3 - x_{6}) - y_3~~ mod ~~p
\end{align*}

Substituting the value of $y_3$, $x_3$ and $x_{6}$,

$y_{6} = \frac{W_6}{U_6} (\frac{W_3^2+x_1U_{3}^2-q_{3}^2(3x_{1}^2+a)^2}{U_{3}^2} - \frac{W_6^2 - 2q_{6}^2(W_3^2+x_1U_{3}^2-q_{3}^2(3x_{1}^2+a)^2)}{U_{6}^2}) - \frac{U_3W_3 (x_1 - x_3) - y_1U_{3}^2}{U_{3}^2}~~ mod ~~p$

Multiplying the equation with $U_{6}^4$,

$U_{6}^4~y_{6} = W_6U_6(q_{6}^2(W_3^2+x_1U_{3}^2-q_{3}^2(3x_{1}^2+a)^2) - (W_6^2 - 2q_{6}^2(W_3^2+x_1U_{3}^2-q_{3}^2(3x_{1}^2+a)^2))) - U_{3}^2q_{6}^4(U_3W_3(x_1 - x_3)-y_1U_{3}^2)~~ mod ~~p$

$U_{6}^4~y_{6} = W_6U_6(q_{6}^2W_3^2+x_1U_{3}^2q_{6}^2-q_{6}^2q_{3}^2(3x_{1}^2+a)^2-W_6^2+2q_{6}^2W_3^2+2q_{6}^2x_1U_{3}^2-2q_{6}^2q_{3}^2(3x_{1}^2+a)^2) - U_{3}^2q_{6}^4(U_3W_3(x_1 - x_3)-y_1U_{3}^2)~~ mod ~~p$

$U_{6}^4~y_{6} = W_6U_6(3q_{6}^2W_3^2+3x_1U_{3}^2q_{6}^2-3q_{6}^2q_{3}^2(3x_{1}^2+a)^2-W_6^2) - U_{3}^2q_{6}^4(U_3W_3(x_1 - x_3)-y_1U_{3}^2)~~ mod ~~p$

Substituting $x_3$ value in equation \ref{eq:mc2.28},

$U_{6}^4~y_{6} = W_6U_6(3q_{6}^2W_3^2+3x_1U_{3}^2q_{6}^2-3q_{6}^2q_{3}^2(3x_{1}^2+a)^2-W_6^2) - U_{3}^2q_{6}^4(U_3W_3(x_1 - \frac{W_3^2+x_1U_{3}^2-q_{3}^2(3x_{1}^2+a)^2}{U_{3}^2})-y_1U_{3}^2)~~ mod ~~p$

$U_{6}^4~y_{6} = W_6U_6(3q_{6}^2W_3^2+3x_1U_{3}^2q_{6}^2-3q_{6}^2q_{3}^2(3x_{1}^2+a)^2-W_6^2) - U_{3}q_{6}^4(W_3(U_{3}^2x_1 -W_{3}^2-x_1U_{3}^2+q_{3}^2(3x_{1}^2+a)^2)-y_1U_{3}^3)~~ mod ~~p$

$U_{6}^4~y_{6} = W_6U_6(3q_{6}^2W_3^2+3x_1U_{3}^2q_{6}^2-3q_{6}^2q_{3}^2(3x_{1}^2+a)^2-W_6^2) - U_{3}q_{6}^4(W_3q_{3}^2(3x_{1}^2+a)^2-W_{3}^3-y_1U_{3}^3)~~ mod ~~p$
\begin{equation}
y_{6} = \frac{N{x6}}{U_{6}^4}~~ mod ~~p
\end{equation}

\subsection{Another Implementation of 10Q}\label{sec:10q}
Here is another example of different implementation of 10Q that was computed previously in Section~\ref{sec:plus2p} through~$2^3 + 2Q$ form. In this section, we will implement 10Q by doubling the point 5Q which were computed previously by applying a differential addition to the algorithm 4Q in Section~\ref{sec:da}. We have succeeded to prove that we can compute 10Q with a single inverse as well.

As we have done in the previous section, we will substitute the $x_5$ and $y_5$ coordinates to the doubling slope in Equation~\ref{eq:mc2.1},
\begin{align*}
\lambda_{10} = \left(\frac{3x_{5}^2+a}{2y_5}\right)~~ mod ~~p 
\end{align*}
\begin{align*}
\lambda_{10} = \frac{3(\frac{(N_y-y_1U^3)^2-x_1U^2(N_x-x_1U^2)^2-N_x(N_x-x_1U^2)^2}{U^2(N_x-x_1U^2)^2})^2+a}{2(\frac{U(N_x-x_1U^2)((N_y-y_1U^3)x_1-(N_y-y_1U^3)x_5- U(N_x-x_1U^2)y_1)}{U^2(N_x-x_1U^2)^2})}~~ mod ~~p 
\end{align*}
Multiplying with $\frac{U^4(N_x-x_1U^2)^4}{U^4(N_x-x_1U^2)^4}$
\begin{align*}
\lambda_{10} = \frac{3((N_y-y_1U^3)^2-x_1U^2(N_x-x_1U^2)^2-N_x(N_x-x_1U^2)^2)^2+aU^4(N_x-x_1U^2)^4}{2U^3(N_x-x_1U^2)^3((N_y-y_1U^3)x_1-(N_y-y_1U^3)x_5- U(N_x-x_1U^2)y_1)}~~ mod ~~p 
\end{align*}
Reforming $x_5$ as,
\begin{align*}
x_{5} = \frac{N_{x5}}{U^2(N_x-x_1U^2)^2}~~ mod ~~p 
\end{align*} 
Then, substitute $x_5$ in the slope equation,

$\lambda_{10} = \frac{3((N_y-y_1U^3)^2-x_1U^2(N_x-x_1U^2)^2-N_x(N_x-x_1U^2)^2)^2+aU^4(N_x-x_1U^2)^4}{2x_1U^3(N_x-x_1U^2)^3(N_y-y_1U^3)-2U(N_x-x_1U^2)(N_y-y_1U^3)N_{x5}- 2U^4(N_x-x_1U^2)^4y_1)}~~ mod ~~p$

Take out a common factor $U(N_x-x_1U^2)$ of the denominator,

$\lambda_{10} = \frac{3((N_y-y_1U^3)^2-x_1U^2(N_x-x_1U^2)^2-N_x(N_x-x_1U^2)^2)^2+aU^4(N_x-x_1U^2)^4}{U(N_x-x_1U^2)(2x_1U^2(N_x-x_1U^2)^2(N_y-y_1U^3)-2N_{x5}(N_y-y_1U^3)- 2y_1U^3(N_x-x_1U^2)^3)}~~ mod ~~p$

Considering,
\begin{equation}
\lambda_{10} = \frac{W_{10}}{U_{10}}~~ mod ~~p
\end{equation}

Where,
\begin{equation}
U_{10} = U(N_x-x_1U^2)q_{10}~~ mod ~~p
\end{equation}

Now we substitute $\lambda_{10}$ in x and y coordinates equations,
\begin{align*}
x_{10} = \lambda_{10}^2 - 2x_5~~ mod ~~p
\end{align*}
\begin{align*}
x_{10} = \left(\frac{W_{10}}{U_{10}}\right)^2 - 2\left(\frac{N_{x5}}{U^2(N_x-x_1U^2)^2}\right)~~ mod ~~p
\end{align*}

Multiplying the equation with $U_{10}^2$,
\begin{align*}
U_{10}^2~x_{10} = W_{10}^2 - 2q_{10}^2N_{x5}~~ mod ~~p
\end{align*}
\begin{equation}
x_{10} = \frac{W_{10}^2 - 2q_{10}^2N_{x5}}{U_{10}^2}~~ mod ~~p
\end{equation}

Now we find $y_{10}$,
\begin{align*}
y_{10} = \lambda_{10} (x_5 - x_{10}) - y_5~~ mod ~~p
\end{align*}
\begin{align*}
y_{10} = \frac{W_{10}}{U_{10}} \left(\frac{N_{x5}}{U^2(N_x-x_1U^2)^2} - \frac{W_{10}^2 - 2q_{10}^2N_{x5}}{U_{10}^2}\right) - \frac{N_{y5}}{U^2(N_x-x_1U^2)^2}~~ mod ~~p
\end{align*}

Multiplying the equation with $U_{10}^3$,
\begin{align*}
U_{10}^3~y_{10} = W_{10} (N_{x5}q_{10}^2 - W_{10}^2 + 2q_{10}^2N_{x5}) - N_{y5}q_{10}^2U_{10}~~ mod ~~p
\end{align*}
\begin{equation}
y_{10} = \frac{W_{10} (N_{x5}q_{10}^2 - W_{10}^2 + 2q_{10}^2N_{x5}) - N_{y5}q_{10}^2U_{10}}{U_{10}^3}~~ mod ~~p
\end{equation}

\subsection{Results}\label{sec:results}
Simulation experiments are performed with a Java implementation of the proposed equations. We have applied the equations on large parameters defined in the standard curve P-521 from the National Institute of Standards and Technology (NIST). Table 2 shows the differences in time and operations count between the new method as compared to the original equations.

As noted in the Table~\ref{fig:mf1}, there is a significant improvement in the speed for our algorithms as compared to the original ones.
The improvement achievable with parallelization can also be observed, together with its scalability for higher orders multiplication scalars.
 
\begin{table}[!th]
\begin{center}
\begin{tabular}{|l|c|c|c|c|l|}\hline
{\bf Routine} & \multicolumn{4}{c|}{\bf Number of Operations} & {\bf Time ms}  \\ \cline{2-5}
  & Mults & Divs & ALUs & Parallel levels &    \\  \hline
4P Eq. & 350/642 & 291/628 & 309/634 & 310/646 & 0.5/0.9 \\ \hline
8P Eq. & 405/930 & 287/909  & 320/921 &  313/939 & 0.6/1.1  \\ \hline
16P Eq. & 518/1248 & 289/1220  &  347/1236 & 327/1260 & 0.9/1.6  \\ \hline
\end{tabular}
\end{center}
\caption{Algorithms Preliminary Measurements.}\label{fig:mf1}
\end{table}  

\section{Sample Applications}
In Section~\ref{sec:mn}, we have seen how to compute a scalar multiple k for elliptic curve point, with a single inverse. Furthermore, we succeeded to perform a differential addition between two points where one of them can be doubled up to 4 times in one equation with a single inverse, as well. This optimization can therefore speed up the computation of many applications and algorithms that are based on Elliptic Curves. In addition, it has the potential to play an important role in developing post quantum cryptosystems such as SIDH. In this section we clarify some of these applications to SIDH.   

At each stage of the Super Singular Isogeny Diffie-Hellman (SIDH) protocol, the kernel of an isogeny has to be computed by both Alice and Bob, calculating the equation P + [k]Q, where P and Q are points on the curve and k is the secret key that is generated by both of them~\cite{jao2011towards}. This operation must be performed in both phases of SIDH. In the first key generation phase, the point is known in advance. In this case, one can construct a look up table that contains all doubles of point Q and reuse any of them as needed. 
   
As shown in Section~\ref{sec:mn}, the direct equations for finding a higher doubling order avoids the original steps. Furthermore, our implementation has proved that the new equations are faster than the original ones, as described in Section~\ref{sec:results}. The introduced optimization speeds up the elliptic curve computation and is beneficial for SIDH's 
quantum security margin.
The new equations can be applied to the right-to-left fast exponentiation algorithm
for binary elliptic curves~\cite{joye2007highly,oliveira2014fast}, 
also used accordingly for Montgomery curves~\cite{oliveira2017pre,faz2018faster}. In addition, it can also be applied to the left-to-right fast Double-and-Add or Double-Add-\&-Subtract algorithms.
We modify these algorithms in order to exploit the new equations.  

\section{Algorithms Overview}

\begin{figure}[!th]
\begin{center}
\epsfig{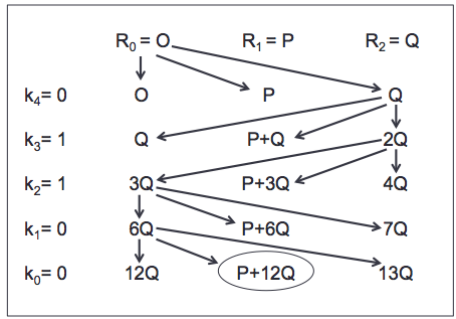}
\end{center}
\caption{Three point ladder (left-to-right)~\cite{jao2011towards,faz2018faster}.}\label{fig:mf}
\end{figure}

This section introduces new algorithms for computing $P + [k]Q$.
The Three-point ladder algorithm (left-to-right or Double-and-Add)~\cite{jao2011towards}, is shown in Figure~\ref{fig:mf}. The example, uses the same scalar the author assumed, which is 12. In order to apply the new equations, one can first compute the tripling equations, then input the result to the $2^{rd}$ double equation. On the other hand, expanding on the right-to-left fast multiplication algorithm, one computes the $2^{nd}$ double, then further applies a single double to the result, and have the results summed. Thus, the right-to-left algorithm is slower than the reverse in this case since it needs three steps while left-to-right algorithm needs two. 

In addition, both algorithms use two accumulators to have intermediary results added to the third accumulator which stores the expected output. In contrast, our equations can shorten most of the steps the processor consumes in calculating the final result by using a single accumulator. 

\begin{figure}[!th]
\begin{center}
\epsfig{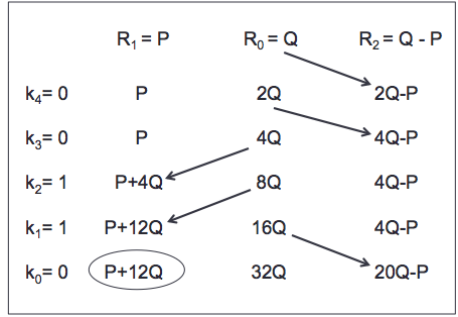}
\end{center}
\caption{Right-to-left algorithm~\cite{faz2018faster}.}
\label{fig:of}
\end{figure}

The right-to-left algorithm~\cite{faz2018faster,joye2007highly,oliveira2014fast,faz2018faster} in Figure~\ref{fig:of} lets the processor calculate the double at each step, regardless of the value of the secret key bit. With the new equations to compute repeated doubles, one no longer needs to double at every step, enabling the development of more efficient algorithms introduced in the next section.  

\section{Fast Multiplication with Mixed Base Multiplicands}
Here follows the description of a few algorithms that can integrate the fast repeated doubling techniques mentioned so far by applying mixed base multiplicands.
With the algorithm $mP + nQ$ one can compute multiplications with scalars up to 31. One can divide $m$'s binary representation into blocks of five bits.
In case an obtained block represents one of the unimplemented scalar multiplications, such blocks may be reduced in length.

\begin{figure}[!th]
	\begin{center}
		\epsfig{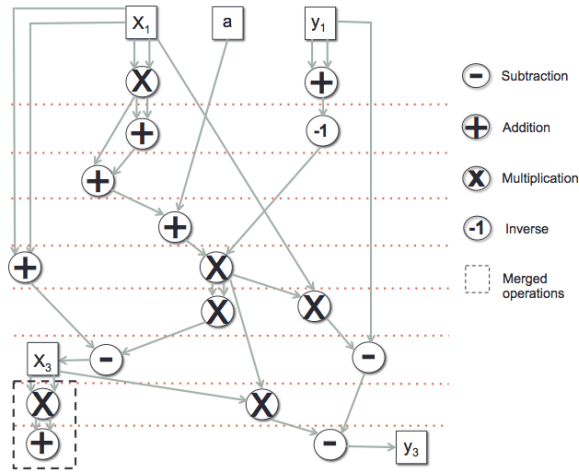}
	\end{center}
	\caption{Data-dependency graph for calculating a single double merged with another one (Parallelization characteristic).}
	\label{fig:if}
\end{figure}

\subsection{Right-to-left Extensions}
The right-to-left algorithm computes the $k[Q]$ by scanning the bits of the scalar $k$ from right to left. It accelerates the computation in the SIDH key generation phase while the left-to-right algorithm increases efficiency in the key exchange phase~\cite{faz2018faster}. 

Here we present an improved right-to-left algorithm that computes more efficiently the kernel of isogenies 
$P + [k]Q$ in both phases. Unlike the Three ladder and right-to-left algorithms, this algorithm scans till it finds the first 1 bit of the secret key $k$, then starts its main loop. Once a 1 bit is located, it applies the improved equations to compute the double that matches the binary order of that bit. For example, if the first one bit was found at the third position from right, one applies the $2^{nd}$ double equations.

\begin{figure}[!th]
	\begin{center}
		\epsfig{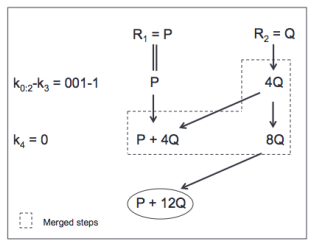}
	\end{center}
	\caption{First Proposed Algorithm.}
	\label{fig:wf}
\end{figure}

One can take advantages of the parallelization characteristic 
offered by the new equations. Namely, assuming that the $x$ coordinate is always computed before $y$, then one can start steps from the next doubling operation before the end of the current one. This saves at least $M+1$ clock cycles (see Figure~\ref{fig:if}). Moreover, mixed addition provides more efficiency to addition and doubling elliptic curve algorithms, especially for large parameters~\cite{bernstein2007faster}. 
Figure~\ref{fig:wf} shows the steps for calculating $P + [k]Q$ for our algorithm when the scalar is a 5-bit number $(12)_{10} = (01100)_{2}$. This parallelization characteristic can be applied to all algorithms described further. 

\begin{algorithm}[!ht]
\caption{Right-to-left Extensions}\label{alg:das1}
\Proc{MultiplyR2L(k, P)} {
 R := O\;
 H := P\;
 \lIf{$|k|>0$ and $k_0==1$}{R = H\;}
 \For{(int i := 1; $i<|k|$; )} 
 {
 	\lnl{line:l2}l := 0\;
 	\Do{($k_i==0$ and $l < 4$)} {
 		$l ++$\;
 		$i ++$\;
 	}
    \lnl{line:l1} H := $2^l H$\;
    \lnl{line:l3}\lIf {($k_i$)} {$\|parallel(R := R + H)\|$\;}
 }
 \lnl{line:l4} \Return $R$\;
}
\end{algorithm}
\begin{algorithm}[!ht]
	\caption{Right-to-Left Knapsack}\label{alg:das2}
	\Proc{MultiplyR2LKnapsack(k, P)} {
		R := O\;
		H := P\;
		\lIf{$|k|>0$ and $k_0==1$}{R := H\;}
		\For{(i := 1; $i<|k|$; )} 
		{
			\lnl{line:a2:l1}l := 0\;
			\Do{($k_i==0$)} {
				$l ++$\;
				$i ++$\;
			}
			\lnl{line:a2:l2} H := DoubleKnapsack($H$, $l$)\;
			\lnl{line:a2:l3}
			{$\|parallel(R := R + H)\|$\;}
		}
		\lnl{line:a2:l4} \Return $R$\;
	}
\end{algorithm}

In Algorithm~\ref{alg:das2} the Right-to-Left extension is further generalized to support dynamic implementation of the doubling to high orders, as a combination of available implementations of lower order that provide the optimal combined performance (at Line~\ref{line:a2:l2}).

In Line~\ref{line:a2:l1} of the pseudocode,  the counter l is initialized  with value zero. 
We scan for the next 1-bit to detect whether there is a following double to be launched in parallel, once the computations of the x and y coordinates start. In Line~\ref{line:a2:l2} the repeated doubling is applied to the point to the order that is specified by the counter l.

In Algorithm~\ref{alg:das1} at Line~\ref{line:l2} this is performed directly
from one of the new doubling equations, while in Algorithm~\ref{alg:das2} at Line~\ref{line:a2:l2} one uses a knapsack optimization to select a combination of available repeated doubling techniques that
best together produce multiplication with $2^l$ (this being the main difference between the two algorithms).

As we see in Line~\ref{line:a2:l3}, the algorithm provides the
opportunity to parallelize aforementioned independent computations.
Finally, in Line~\ref{line:a2:l4}, the register R will have the final result.

\subsection{Double and Add Extensions}     
\begin{figure}[!th]
\begin{center}
\epsfig{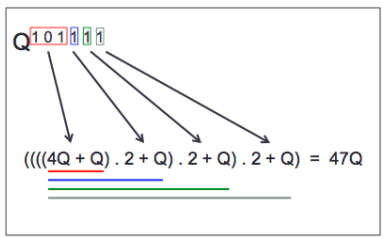}
\end{center}
\caption{Left-to-Right Proposed Algorithm.}
\label{fig:lr}
\end{figure}

In Section~\ref{sec:mn} 
it is shown how to
compute all intermediate exponent and mix doubling with a differential addition with a single inverse.
The left-to-right algorithm
starts scanning from left the next one-bit considering that the most significant bit is one. Then, it decides weather it applies doubling or doubling and addition depending on the data being read. For instance, if the first two one bits were representing the binary equivalent $(101)_2$ which is $5_{10}$, it will multiply the base by 4 because it was shifted to the left by two bits.
Since the last bit scanned is a 1, it also applies a differential addition to the point being doubled with the base point. Thus, the implementation will be $4Q+Q$. Figure~\ref{fig:lr} shows a practical example for calculating $Q^{47}$.
This technique computes $Q^{47}$ with only four inverses, instead of the eight inverses when performing the original equations.

\begin{algorithm}
	\caption{Double and Add Extensions}\label{alg:das3}
\Proc{MultiplyL2RKnapsack(k, P)} {
	\lIf{$|k|\leq 0$ or $k_{|k|-1}==0$}{return O\;}
	D := P\;
	\For{(int i := $|k|-2$; $i\geq 0$; )} 
	{
		l := 0\;
		\Do{($k_i == 0$ and $i \geq 0$)}{
			i - -\;
			l ++\;
		}
		\eIf{$k_i == 1$} {
			\lnl{line:L2R:l4} D := DoubleAndAddKnapsack(D, l, P)\;
		} {
			\lnl{line:L2R:l5} D := DoubleKnapsack(D, l)\;
		}
	}
	\Return D\;
}
\end{algorithm}

In Line~\ref{line:L2R:l4} of the pseudocode of the Double-and-Add extensions we apply the DoubleAndAddKnapsack function taking
as parameter the counter l that specifies the current bit location, and the base point P to be added at the end. 
Otherwise, we apply the DoubleKnapsack function that computes the shifting to the left by multiplying the D value with $2^l$, which is the same as the function used in Algorithm~\ref{alg:das2}.
   
\subsection{NAF Extensions}
\begin{figure}[!th]
\begin{center}
\epsfig{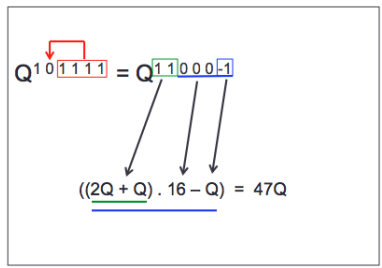}
\end{center}
\caption{Left-to-Right Double-Add-and-Subtract Algorithm.}
\label{fig:as}
\end{figure}

The last proposed algorithm is a version of the Left-to-Right algorithm called Double-Add-and-Subtract. The original Double-Add-and-Subtract minimizes the number of 1s in the binary representation of the scalar by repeatedly detecting sequences of $k$ 1-bit at position $(i)$ to $(i+k-1)$ and replacing them with $2^{i+k}-2^i$. The initialization of this technique minimizes the Hamming weight of a scalar multiplication by using the non-adjacent form (NAF). Likely, with a random scalar multiplier, half of the bits in its original binary representation will be non-zero. However, with NAF's special representation this will be dropped to one third.
For example,~$15_{10} = 1111_{2}$. This is rewritten as~$(16-1)_{10} = \langle 1,0,0,0,{-1}\rangle$, where sequences $\langle b_k,...,b_0\rangle$ are interpreted as $\sum_{i=0}^k 2^ib_i$ for $b_i\in\Z$, most commonly used with $b_i\in\{-1,0,1\}$. 
The Left-to-Right Double-Add-and-Subtract Algorithm in the past could not improve upon three leading 1s, but the new efficient 
$8P$, 
algorithm can improve speed for sequences of 3 leading 1s.
One can predict automatically the fastest combination of repeated doubling operations. Figure~\ref{fig:as} shows the same example, calculating $[47]Q$ by applying the Left-to-Right Double-Add-and-Subtract Algorithm. 
This computes
$[47]Q$ with only two inverses unlike the Left-to-Right Double-and-Add Algorithm that needed 4.

\begin{algorithm}[!ht]
	\caption{NAF Extensions}\label{alg:das4}
	\Proc{MultiplyL2RMix(k, P)} {
		(m,B) := mixed\_NAF\_representation\_Knapsack(k)\;
		D := multiply(P,$m_{|m|-1}$)\;
		\For{(int i := $|m|-2$; $i\geq 0$; )} 
		{
			// $D := D * B_i + P*m_i$\\
			D := RadixPromoteAndAdd(D, $m_i$, $B_i$, P)\;
		}
		\Return D\;
	}
\end{algorithm}

In general, for a number $k$, a mixed base representation $m=\overline{m_N...m_0}$,
with bases $B=\langle B_N,...,B_0\rangle$, must have the
property that 
$$k=\sum_{i=0}^N m_i\prod_{j=0}^{i-1} B_j$$
A canonical mixed base representation expects that $\forall i, 0\leq m_i < B_i$, but with NAF this additional condition is not enforced.

The Algorithm~\ref{alg:das4} starts by recasting $k$ in
a mixed base $B$ with NAF representation $m$,
selected such that $D*b_i + P*m_i$ can be computed efficiently
given available software libraries (implementations of direct repeated doubling operations).
Then, the Multiply Left to Right Mix (MultiplyL2RMix) proceeds one digit at a time,
promoting the current value based on the radix of the corresponding position and adding/subtracting the digit's multiple of the base point $P$.
The {\tt RadixPromoteAndAdd} procedure can use memoisation dynamic programming to store encountered multiples of the base $P$, and reuse them on subsequent calls.

The simplest implementation of {\tt mixed\_NAF\_representation\_Knapsack} is base 16, namely where $\forall i, b_i=16$.
Another alternative where base 32 can be used at times, and base 16 any time, is in Algorithm~\ref{alg:rep}.

\begin{algorithm}[!ht]
	\caption{Mix NAF Representation bases 16 and 32}\label{alg:rep}
	\Proc{mixed\_NAF\_representation\_Knapsack(k)} {
		n := 0\;
		carry := 0\;
		\For{(int i := 0; $i \leq |k|$;)} {
			j = min($|k-1|$, i + 4)\;
			\eIf{(optimizable($\overline{k_j...k_i}$ + carry))} {
				$b_n := 2^5$\;
			} {
			    j - -\;
			    $b_n := 2^4$\;
		    }
			$(m_n, carry) := NAF(b_n,\overline{k_i...k_j} + carry)$\;
	       n ++\;
	       $i := j-1$\;
		}
	   \lIf{$carry > 0$}{$m_n := carry$}
	\Return ($m,b$)
	}
\end{algorithm}
Here procedure NAF reduces the current digit to an optimizable 
positive or negative multiplier, and a carry value.

\section{Fast Multiplication with Base 16 Multiplicands}
Here we mention a simple special case algorithm based on base 16 representations of the multiplicands. 
For example $35=23_{16}$. 
Then, for a multiplier of the form $\overline{qr_{16}}$ the computation is implemented by as,
\begin{align*}
16(qP) + rP ~~ mod~~ p
\end{align*}

For the scalar $10150=27A6_{16}$, the obtained algorithm is equivalent to:

\begin{align*}
((16(2P) + 7P)(16)+ 10P)(16) + 6P ~~ mod~~ p
\end{align*}  

If one precomputes the point 16P, then one can efficiently compute the point 10150P with the total of four inversions by applying the new direct equations for $2^4P+nQ$.

\section{Conclusion}
We have proposed optimized methods for computing scalar multiplication in Elliptic Curves over a prime field, for the short Weierstrass form in the affine plane. We have described a methodology for direct repeated point doubling with high order, as well as point addition of the form $nP + mQ$ by using a single inversion. These new algorithms are shown to be significantly faster than the original equations. In addition, we have developed optimized equations 
for repeated doubling of orders that are higher compared to what is available with related existing algorithms (up to 31). 

In its second part the report introduces new fast elliptic curve multiplication procedures extending current variants and alternatives of Double-and-Add to exploit mixed-base representations of the scalar multiplicand, potentially in non-adjacent form, tuned to exploit available optimized implementations of given scalars.
Such given optimized scalars may either have special hardware support or, for example, be using only a single inverse when working in the affine plane.
Parallelization opportunities are also highlighted. Our empirical implementation indicates that proposed equations outperform the original methods and also provides a significant speed-up when implemented in hardware.

\bibliographystyle{plain}
\bibliography{bibfilethesis}

\end{document}